\documentclass[12pt,reqno]{amsart}
\usepackage{amssymb}
\usepackage{amsfonts}
\usepackage{amsmath}
\usepackage{graphicx}
\usepackage[margin=1.0in]{geometry}
\providecommand{\U}[1]{\protect\rule{.1in}{.1in}}

\newtheorem{theorem}{Theorem}
\theoremstyle{plain}

\newtheorem{corollary}{Corollary}

\numberwithin{equation}{section}
\begin{document}
\title[Kinematic Condition For Solitons]{Kinematic Condition For Soliton Motions of an $n$-dimensional continuum in
$R^{n+m}$}
\author{Namik Ciblak}
\address{Department of Mechanical Engineering\\
Yeditepe University, Istanbul}
\email{nciblak@yeditepe.edu.tr}
\urladdr{http://me.yeditepe.edu.tr}
\date{December 1, 2016}

\begin{abstract}
A new kinematic condition for soliton motions of an $n$-dimensional continuum
in $R^{n+m}$, independent of the underlying physics, is proven. The condition
and its consequences for different cases are demonstrated. A soliton in a 1D
string that rocks back and forth, a rotating soliton in a 2D membrane, and
various other cases are presented as examples. It is shown that traveling
knots based on classical wave equation are plausible. Cases in which all the
motions are solitons are also presented. Compatibility of equations of motion
with the kinematic constraint is explored and demonstrated.

\end{abstract}
\maketitle

\section{Introduction}

All physical phenomena are believed to obey some certain rules, known or yet
to be discovered, that are potentially expressible in mathematical forms. In
the experience of scientific exploration, a dominating mathematical
representation of such phenomena turned out to be partial differential
equations (PDE). Owing to the origins of PDE, these are usually called
\emph{the equations of motion}, despite the fact that they may actually
describe the dynamics of quantities quite different from "motion proper", such
as temperature, pressure, or some other field.

Nevertheless, one of the earliest PDEs was the classical wave equation
describing the actual motion of a string as proposed by D'Alembert in 1747
\cite{Dalembert1}. This is not surprising since the most readily observable
dynamics of physical phenomena concern the position, especially so in the
beginning of the scientific era. In order to stress this point, we must note
that, what nowadays is an introductory example in theory of PDEs, the dynamics
of a simple string sparked a dispute that took more than a century to settle.
The story of this controversy can be found in two excellent reviews, one by
Wheeler and Crummett (1987) \cite{Wheeler}, the other by Zeeman (1993)
\cite{Zeeman}. In somewhat similar way, this study evolved out of the author's
involvement in research on the dynamics of strings.

Although the origin of the problem involved 1D strings, this study is
concerned with motions of deformable material bodies in general. In the
context of PDEs, any topological space with Hausdorff property, which makes
the continuum assumption plausible, can be taken as a representation of a
material body, provided that it is assigned a metric structure. In doing so,
all the tools of the continuum mechanics become available. Here, we take the
simplest of such spaces, namely the $n$ dimensional Euclidean space $R^{n}$.
We disregard any boundaries and take whole of $R^{n}$ as our material body. By
\textit{motion of a material body,} we mean any allowable 1-parameter map of
$R^{n}$ into $R^{n+m}$, where the parameter is time, such as that of a 1D
string in plane motion: a map from $R$ into $R^{2}$.

Although no specific set of equations of motion are required, most examples
would naturally come from those encountered in the theory of elasticity or in
the theory of fluids -- continuum mechanics in general. Nevertheless,
continuous models of any other physical or chemical phenomena may also provide
examples, of course.

Among special motions that can be observed in such systems, solitons are of
special importance. This is mainly due to the fact that they are observed not
only in simple systems, but in many other quite complicated phenomena ranging
from quantum mechanics to optics, and to other advanced physical theories of
matter, even to social dynamics.

Classical solitons are localized motions, or lumps, that preserve their shapes
during propagation. The first known observation of this phenomenon was
reported in 1844 by John Scott Russell, a civil engineer and naval architect,
who, after observing a wave of disturbance in a narrow water channel, wrote of
"\emph{a large solitary elevation, a rounded, smooth and well-defined heap of
water}" that persisted for one or two miles \cite{russell}. Later, a theory
for the motion of water waves in shallow rectangular channel was proposed by
Boussinesq in 1871 in support of Russell's observation \cite{Boussinesq}, who
was seconded by Lord Rayleigh in 1876 \cite{Rayleigh}.

One must note that D'Alembert's solutions to the wave equation,
\cite{Dalembert1}, actually heralded the existence of solitons almost exactly
a hundred years earlier than Russell's report. Only, by allowing any graph
moving with a constant velocity, not necessarily a soliton, D'Alembert's
solutions were too general.

Since these beginnings, solitons became and still are an active area of
research, especially as they pertain to nonlinear partial differential
equations such as Korteweg--de Vries equation \cite{KdV}, nonlinear
Schr\"{o}dinger equation \cite{Zakharov}, Sine-Gordon equation, Burger's
equation, and many others, including many modified versions \cite{PDEhandbook}.

Solitons can be found in many other diverse areas in which the related
processes have some dynamical models. These areas include signals in networks
of any kind, from neural networks to electrical and social networks; motion of
crystal lattice structures; dynamics of social and economic phenomena; and,
evolutionary and biological processes, \cite{LiqCrystal} \cite{biology}
\cite{ANNsoliton}.

Literature on solitons is so vast that any attempt to summarize all would
certainly be futile and unfair. Nevertheless, it seems that the body of
research on solitons is silent on the question of why and how a soliton
solution is viable for a given problem. The main approach seems to be that,
given the equation of motion, the underlying physics, one tries to search for
viable soliton solutions. In this study we show that there is a more stringent
condition which is \emph{independent of the physics of the problem}.

Our focus is on the motions of an $n$-dimensional body in an $\left(
n+m\right)  $-dimensional Euclidean space and we develop a kinematic condition
for the existence of soliton motions. Therefore, any proposed soliton solution
for a particular problem would have to satisfy this kinematic condition first,
regardless of the equations of motion, making it a necessary condition.

In order to demonstrate the result some special cases are presented ranging
from simple cases of transverse and longitudinal motions, to the motion of 1D
string in $(m+1)$ dimensions, to the motion of a 2D membrane in 3D space. As
some interesting cases, we also present variable velocity solitons such as
rotating lumps in 2D membranes, oscillating lumps in 1D strings.

Perhaps, the most striking example involves the plausibility of a soliton knot
based on simple wave equations. Finally, compatibility of the kinematic
condition with the equations of motion is discussed.

Though not attempted here, the results of the present study should also be
extendible to manifolds with more complicated topologies and metrics.

As it is not the goal of this paper to develop a detailed mathematical
analysis of the soliton phenomenon, the mathematical statements are formed
quite loosely. For example, throughout the study we, tacitly or otherwise,
applied continuously differentiable condition to all functions, despite the
fact that results can easily be extended to continuous but piecewise
differentiable cases.

Further, in defining a soliton we sometimes allowed displacements that do not
vanish, or are unbounded, at large distances or times. An example of this is
$x-t$, which, for the purposes of this study, is a legitimate soliton, though
not physically viable or desirable. In actuality, even motions that blow up at
finite distances or times are mathematically admissible, even if not physically.

\section{The Soliton Motion}

Let $\Xi$ be a deformable, continuous body, the reference state $\Xi_{0}$ of
which is represented by an $n$ dimensional Euclidean manifold, $R^{n}$.
Further, let the points of the body be allowed to move in a larger manifold,
$R^{n+m}=R^{n}\times R^{m}$. We consider a coordinate system in $R^{n}$ so
that the position vectors of material points of $\Xi$ in the reference state
$\Xi_{0}$ are denoted by $\bar{x}=\left[  x_{1},x_{2},\ldots,x_{n}\right]
^{T}$, where $x_{i}$ are the coordinates.

In order to describe the motion we consider the displacement vectors $\bar
{u}\left(  \bar{x},t\right)  $ and $\bar{v}\left(  \bar{x},t\right)  $, in
$R^{n}$ and $R^{m}$, respectively, of a material point referred to by
coordinates $\bar{x}$ in the reference state. Thus, the configuration of $\Xi$
in $R^{n+m}$ at any time $t$ is given by the map $\left(  \bar{x}+\bar
{u}\left(  \bar{x},t\right)  ,\bar{v}\left(  \bar{x},t\right)  \right)  $,
which we shall call as the graph from here on. Each point of the graph can be
referred to by a vector $\bar{r}=\left[  \bar{x}+\bar{u},\bar{v}\right]  ^{T}
$. Here, $v_{i}$ also serve as coordinates for $R^{m}$. We will call $R^{m}$
as the transverse directions with respect to $\Xi_{0}$.

\begin{figure}
[ht]
\begin{center}
\includegraphics[
height=1.9458in,
width=4.0145in
]
{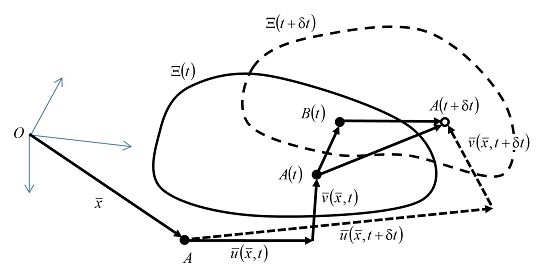}
\caption{A material body in motion as depicted at two instances of time: $t$
and $t+\delta t$. The reference configuration $\Xi_{0}$ at $t=0$ is described
by coordinates $x_{i}$.}
\label{mainFig}
\end{center}
\end{figure}

Figure \ref{mainFig} depicts the positions of a particular point $A$ at time
$t$ and $t+\delta t$, where $\delta t$ denotes an arbitrarily small time
interval. The displacement of $A$ from $t$ to $t+\delta t$ is given by
\begin{equation}
\overline{A\left(  t\right)  A\left(  t+\delta t\right)  }=\frac{\partial
\bar{r}}{\partial t}\delta t=\bar{r}_{t}\delta t=\left[
\begin{array}
[c]{c}
\bar{u}_{t}\\
\bar{v}_{t}
\end{array}
\right]  \delta t
\end{equation}
where $\bar{u}_{t}$ and $\bar{v}_{t}$ are the velocity components in $R^{n}$
and $R^{m}$, respectively.

In classical soliton motion, the shape seems to move in a given direction with
a constant velocity, despite the fact that the actual motion of the particles
could be quite different. We will call this as the apparent motion and its
associated velocity as the apparent velocity or soliton velocity, which
corresponds to the \emph{group velocity} in wave mechanics. This apparent
motion also has the property that the points move in such a way that different
points at different times are seen as the same point moving with the apparent
velocity. An example for this is the point at the crest of a single
disturbance wave in water: the crest seems to move in a certain direction with
a certain velocity, whereas, in actuality, it is formed by different points at
different times.

If the body $\Xi$ is to have a soliton motion in $R^{n+m}$ then the apparent
motion of $A$ would be such that as if $A(t+\delta t)$ originates from another
point $B$ at time $t$, exactly $-\bar{C}\delta t$ distance away, where
$\bar{C}$ is the soliton velocity in $R^{n+m}$. In classical soliton motions
this velocity is restricted to $R^{n}$ and is usually taken as constant. For
now, we allow the most general case of $\bar{C}\left(  \bar{x},t\right)  \in
R^{n+m}$. Later, we constrain $\bar{C}$ to be at most a function of time in
order to comply with basic features of a soliton.

In order to find where exactly $B\left(  t\right)  $ would have to be, we
consider a small neighborhood of $A\left(  t\right)  $ in which we expect to
find $B\left(  t\right)  $, due to continuity. Let $\delta\bar{x}$ be a small
variation around $\bar{x}$. Then, provided that certain kinematical
constraints are obeyed, starting from $A\left(  0\right)  $, a certain
combination of $\delta x_{i}$ would land on a point $B\left(  0\right)  $ in
the reference state, which, at time $t$, would be at $B\left(  t\right)  $
such that
\begin{equation}
\overline{A\left(  t\right)  B\left(  t\right)  }=\left(  \frac{\partial
\bar{r}}{\partial\bar{x}}\right)  \delta\bar{x}=\left[
\begin{array}
[c]{c}
\tilde{I}+\left(  \bar{\nabla}\bar{u}^{T}\right)  ^{T}\\
\left(  \bar{\nabla}\bar{v}^{T}\right)  ^{T}
\end{array}
\right]  \delta\bar{x}
\end{equation}
where $\bar{\nabla}=\left[  \frac{\partial}{\partial x_{j}}\right]  $ is the
gradient operator, and, $\left(  \bar{\nabla}\bar{u}^{T}\right)  ^{T}$ and
$\left(  \bar{\nabla}\bar{v}^{T}\right)  ^{T}$ stand for matrices of first
partial derivatives, or the Jacobians, $u_{i,j}=\left[  \frac{\partial u_{i}
}{\partial x_{j}}\right]  $ and $v_{k,j}=\left[  \frac{\partial v_{k}
}{\partial x_{j}}\right]  $, $i,j=1,...,n$, $k=1,...,m$, and $\tilde{I}$ is
the $n\times n$ identity matrix.

The $\delta x_{i}$ quantities must be such that as to satisfy the triangular
vector loop between $A\left(  t\right)  $, $B\left(  t\right)  $, and
$A\left(  t+\delta t\right)  $, as shown in Figure \ref{mainFig}. That is,
\begin{equation}
\left[
\begin{array}
[c]{c}
\tilde{I}+\left(  \bar{\nabla}\bar{u}^{T}\right)  ^{T}\\
\left(  \bar{\nabla}\bar{v}^{T}\right)  ^{T}
\end{array}
\right]  \delta\bar{x}=\left[
\begin{array}
[c]{c}
\bar{u}_{t}\\
\bar{v}_{t}
\end{array}
\right]  \delta t-\bar{C}\delta t
\end{equation}
In order to simplify, we scale $\delta\bar{x}$ by $\delta t$ such that
$\delta\bar{x}=\bar{\alpha}\delta t$, which is allowable since $\delta t>0$.
Also, let $\bar{C}=\left[  \bar{c},\bar{k}\right]  $, where $\bar{c}$ and
$\bar{k}$ are the velocity components in $\Xi_{0}$ and transverse directions,
respectively. With these, we obtain the following two equations.
\begin{align}
\left[  \tilde{I}+\left(  \bar{\nabla}\bar{u}^{T}\right)  ^{T}\right]
\bar{\alpha}  &  =\bar{u}_{t}-\bar{c}\label{eq1main}\\
\left(  \bar{\nabla}\bar{v}^{T}\right)  ^{T}\bar{\alpha}  &  =\bar{v}_{t}
-\bar{k} \label{eq2main}
\end{align}
Now, if the determinant of $\left[  \tilde{I}+\left(  \bar{\nabla}\bar{u}
^{T}\right)  ^{T}\right]  $ is not zero then, from the first, we get
\begin{equation}
\bar{\alpha}=\left[  \tilde{I}+\left(  \bar{\nabla}\bar{u}^{T}\right)
^{T}\right]  ^{-1}\left(  \bar{u}_{t}-\bar{c}\right)
\end{equation}
which, when used in Equation \ref{eq2main}, yields a general condition for
soliton motions. We summarize this result as follows.

\begin{theorem}
\label{MainTheo}If a continuous body, initially embedded in $R^{n}$ with
coordinates $\bar{x}$, moves in $R^{n+m}$ with a soliton velocity of $\bar
{C}\left(  \bar{x},t\right)  $, then its displacement functions $\bar
{u}\left(  \bar{x},t\right)  $ and $\bar{v}\left(  \bar{x},t\right)  $ satisfy
\begin{equation}
\bar{v}_{t}-\bar{k}=\left[  \bar{\nabla}\bar{v}^{T}\right]  ^{T}\left[
\tilde{I}+\left(  \bar{\nabla}\bar{u}^{T}\right)  ^{T}\right]  ^{-1}\left(
\bar{u}_{t}-\bar{c}\right)  \label{eqMain}
\end{equation}
where $\bar{c}$ and $\bar{k}$ are the components of $\bar{C}$ in $R^{n}$ and
$R^{m}$ subspaces, respectively.
\end{theorem}

If one defines $\bar{V}\left(  x,t\right)  =\bar{v}-\bar{k}t$ and $\bar
{U}=\bar{u}+\bar{x}-\bar{c}t$ then Equation \ref{eqMain} simplifies to
\begin{equation}
\bar{V}_{t}=\left[  \bar{\nabla}\bar{V}^{T}\right]  ^{T}\left(  \bar{\nabla
}\bar{U}^{T}\right)  ^{-T}\bar{U}_{t} \label{eqMainSimple}
\end{equation}

The necessity of the existence of the inverse explains the origin of the
aforementioned kinematical constraints. In order to understand the nature of
this restriction one can look at the one-dimensional case, in which the matrix
term reduces to $\frac{1}{1+u_{x}}$. If $u_{x}=-1$ at an isolated point, then
it can be excluded from the analysis and the above condition is still
applicable elsewhere. However, if $u_{x}=-1$ in an open neighborhood of $x$
then $u=-x+f\left(  t\right)  $ in there, which means $x+u=f\left(  t\right)
$ and the whole neighborhood is mapped to a single point $f\left(  t\right)
$. This is rejected since it destroys the integrity of the body. For a real
body this would amount to a violation of conservation principles.

From a mathematical point of view, such pathological maps cause changes in
local or global topological properties, such as dimensionality. Therefore, we
insist on the existence of the inverse except at some isolated points.
Resolution of the case in which inverse fails at some isolated points is left
outside the scope of this study.

In summary, regardless of the underlying physics, any soliton motion must obey
this rule because it is simply based on the definition of a soliton. Some
require more from a soliton, such as maintaining shape even after interactions
with other solitons. This is not followed here since no physical laws are specified.

One has to be careful here: given any continuously differentiable functions
$\bar{u}$ and $\bar{v}$, and any continuous function $\bar{c}$, one can
uniquely determine a $\bar{k}$ such that the kinematic condition is satisfied.
Do these things qualify as proper solitons? Although the answer depends on
individual perspectives, in the scope of this study this question is left as a
matter of definition. In the sequel, we demonstrate some examples in which
this relaxed definition yields interesting cases. Other than these, we usually
revert back to the conventional definition in which $\bar{c}$ is constant and
$\bar{k}$ is zero.

The converse of Theorem \ref{MainTheo} can only be given locally by following
the steps backwards until the triangular vector loop is obtained. This implies
that in a neighborhood of $\bar{x}$ the graph behaves like a soliton. However,
the result cannot be extended to the whole domain since, due to the dependence
of $\bar{C}$ on $\bar{x}$, the velocity of the locally soliton-like motions
will have a spatial variation, in general. This, in turn, means that the shape
of the graph will evolve. We give examples of this sort in the sequel. This
generality is not necessarily undesirable as it may unveil many interesting
phenomena such as evolving solitons, oscillating solitons, and so on. However,
we may specialize it further by constraining $\bar{C}$ to be a function of
time, at most. Then, the soliton motion would be the same everywhere, which
preserves the shape. Hence, we have the following result.

\begin{corollary}
\label{MainTheo2}A continuous body, initially embedded in $R^{n}$ with
coordinates $\bar{x}$, moves in $R^{n+m}$ with a soliton velocity of $\bar
{C}\left(  t\right)  $ if and only if its displacement functions $\bar
{u}\left(  \bar{x},t\right)  $ and $\bar{v}\left(  \bar{x},t\right)  $ satisfy
\begin{equation}
\bar{v}_{t}-\bar{k}=\left[  \bar{\nabla}\bar{v}^{T}\right]  ^{T}\left[
\tilde{I}+\left(  \bar{\nabla}\bar{u}^{T}\right)  ^{T}\right]  ^{-1}\left(
\bar{u}_{t}-\bar{c}\right)  \label{eqMain2}
\end{equation}
where $\bar{c}\left(  t\right)  $ and $\bar{k}\left(  t\right)  $ are the
components of $\bar{C}$ in $R^{n}$ and $R^{m}$ subspaces, respectively.
\end{corollary}

Theorem \ref{MainTheo} and Corollary \ref{MainTheo2} are the main results of
this study.

\section{Special Cases}

The general kinematic conditions, Theorem \ref{MainTheo} and Corollary
\ref{MainTheo2}, for soliton motions do not reveal much at first. In order to
understand the underlying mechanism and implications, some special cases are
investigated in this section.

\subsection{Transverse Wave Solitons}

In this case the solitons belong to the purely transverse waves class in which
the material points only move in transverse directions whereas the soliton
motion is completely in $R^{n}$ directions. This means $\bar{u}=\bar{0}$ and
$\bar{k}=\bar{0}$, and the kinematic condition reduces to following well-known
case.
\begin{align}
\bar{v}_{t}  &  =-\left[  \bar{\nabla}\bar{v}^{T}\right]  ^{T}\bar{c}\\
v_{j,t}  &  =-c_{i}\left(  \bar{x},t\right)  v_{j,x_{i}}
\end{align}
where $i=1,...,n$ and $j=1,...,m$, and summation over $i$ is implied. This is
the kinematic condition for a vector soliton with variable velocity. A special
case is when $c_{i}$ is a function of time only. In such a case, let
$D_{i}\left(  t\right)  $ such that $c_{i}=\frac{dD_{i}}{dt}$. Then, the
solution is
\begin{equation}
\bar{v}=\bar{f}\left(  \bar{x}-\bar{D}\left(  t\right)  \right)
\end{equation}

An example of a soliton with a variable velocity, a 2D membrane moving in 3D,
is presented in the sequel. Also note that, if it happens that the soliton
velocity $c\left(  t\right)  $ is in the direction of a particular coordinate
$x_{I}$, which is always achievable by suitably selecting coordinates, then
\begin{align}
\frac{\partial v_{j}}{\partial t}  &  =-c\frac{\partial v_{j}}{\partial x_{I}
}\ \ \left(  j=1,...,m\right) \\
\bar{v}_{t}  &  =-c\bar{v}_{x_{I}}
\end{align}

In general, given $c_{i}\left(  \bar{x},t\right)  $, one can use the method of
characteristics to determine $v\left(  \bar{x},t\right)  $. It is easy to
demonstrate what happens if the soliton velocity is dependent on spatial
variables. Consider, for example, the 1D case $v_{t}=-c\left(  x\right)
v_{x}$ with $c=\frac{1}{1+x^{2}}$, a quickly decreasing speed. The solution is
$v=f\left(  x+\frac{1}{3}x^{3}-t\right)  $, where $f$ is any continuously
differentiable function. For example, the graph $\left(  x,e^{-\left(
x+\frac{1}{3}x^{3}-t\right)  ^{2}}\right)  $ is initially bell-shaped and
moves like a soliton, yet gradually changes its shape and slows down
dramatically. In this case, the soliton-like motion happens only locally.

Using this general velocity case one can create quite interesting,
soliton-like motions such as evolving solitons. However, our main interest in
this study is in soliton motions in which the shape is preserved at all times.
Therefore, in the sequel we shall only consider cases in which the soliton
velocity is either constant or a function of time.

Nevertheless, the general condition does not always destroy the soliton
character. An example of this is given later in which a 2D membrane executes a
soliton motion with a velocity that depends on spatial coordinates.

\subsection{Longitudinal Wave Solitons}

In this case the solitons are longitudinal waves in which both the material
points and the soliton motion are completely constrained into $R^{n}$. This
means $\bar{v}=\bar{0}$ and $\bar{k}=\bar{0}$. Now, however, the kinematic
condition reduces to a zero identity, providing a null result. Therefore, one
must return to the original conditions, one of which (Equation \ref{eq1main})
becomes, after defining $\bar{u}=\bar{U}-\bar{x}+\bar{c}t$
\begin{equation}
\left(  \bar{\nabla}\bar{U}^{T}\right)  ^{T}\bar{\alpha}=\bar{U}_{t}
\end{equation}
Hence, $\bar{U}=\bar{f}\left(  \bar{x}+\bar{\alpha}t\right)  $ is a soliton
with a velocity of $-\bar{\alpha}$, and
\[
\bar{u}=-\left(  \bar{x}-\bar{c}t\right)  +\bar{f}\left(  \bar{x}+\bar{\alpha
}t\right)
\]
In order to see the behavior of the deformations, one can look at the graph
$\left(  \bar{x},\bar{u}\right)  $, which is a soliton if and only if
$\bar{\alpha}=-\bar{c}$. Therefore, $\bar{u}=\bar{g}\left(  \bar{x}-\bar
{c}t\right)  $.

From a different perspective, one should notice that $\bar{u}\left(
x,t\right)  $ and $\bar{v}\left(  x,t\right)  $ are only functions ascribed to
the points in the reference state. They could very well be quantities other
than displacements. For example, one could associate a single function with
the material points signifying temperature, pressure, or higher order
quantities. In such situations, one would imagine plotting the ascribed
function, with a certain scaling, in a space orthogonal to $R^{n}$, virtually
making them identical to $v\left(  x,t\right)  $. For example, this is how we
plot a pressure wave along a spatial dimension.

Therefore, considering $\bar{u}\left(  x,t\right)  $ as plotted in transverse
dimensions versus $\bar{x}$, a longitudinal wave can be represented by a
transverse wave, which is effectively equivalent to substitutions: $\bar
{u}\rightarrow\bar{v}$, $\bar{0}\rightarrow\bar{u}$, and $\bar{0}
\rightarrow\bar{k}$ in the kinematic constraint. Hence, the resulting
condition is
\begin{align}
\bar{u}_{t}  &  =-\left[  \bar{\nabla}\bar{u}^{T}\right]  ^{T}\bar{c}\\
u_{j,t}  &  =-c_{i}u_{j,x_{i}}
\end{align}
where $i,j=1,...,n$ and summation over $i$ is implied. This is essentially the
same as that for transverse waves. Note that in this case the "shape" of the
soliton is represented by $u\left(  x,t\right)  $.

\subsection{1D String in 2D Motion}

This is a one-dimensional body moving in two dimensions: $n=m=1$. A good
example is the planar motion of an ideal elastic string. In this case, the
soliton condition reduces to
\begin{align}
v_{t}-k &  =\frac{v_{x}}{1+u_{x}}\left(  u_{t}-c\right)  \label{1D2Deq}\\
v_{t}+cv_{x} &  =v_{x}u_{t}-u_{x}v_{t}+k\left(  1+u_{x}\right)
\end{align}
For $k=0$, the condition is
\begin{equation}
v_{t}+cv_{x}=v_{x}u_{t}-u_{x}v_{t}
\end{equation}

For constant $c$ and $k$, it is not difficult to show that the general
solution is
\begin{equation}
u=-x+ct+f(v-kt)
\end{equation}
where $f$ is an arbitrary and continuously differentiable function. For more
general case in which $c$ and $k$ are functions of time only, one would have
\begin{equation}
u=-x+D(t)+f(v-K(t))
\end{equation}
where $\frac{dD}{dt}=c$ and $\frac{dK}{dt}=k$.

For a special case in which $k=0$, $v=f\left(  x\right)  $, such that
$v_{x}\neq0$, the solution would be $u_{t}=c$, or $u=g\left(  x\right)  +ct$.
The graph becomes $\left(  x+g\left(  x\right)  +ct,f\left(  x\right)
\right)  $. One can explicitly show that this graph is a soliton if $x+g$ is
invertible. Let $G\left(  x\right)  =x+g$ be invertible. Then, define
$p=G\left(  x\right)  +ct$, giving $x=G^{-1}\left(  p-ct\right)  $. Hence, the
graph becomes $\left(  p,f\left(  G^{-1}\left(  p-ct\right)  \right)  \right)
$ which is a soliton. Usefulness of Corollary \ref{MainTheo2} becomes obvious
if one consider the cases in which $G\left(  x\right)  $ is not invertible, at
least not explicitly.

Other special cases can also be demonstrated. However, they are outside the
scope of this study. The point here is that non-soliton displacement functions
can give rise to soliton graphs.

\begin{figure}
[ht]
\begin{center}
\fbox{\includegraphics[
height=2.0141in,
width=3.0208in
]
{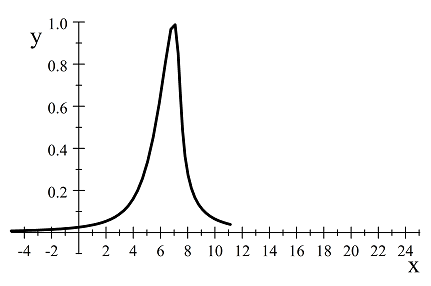}
}\caption{Graph of $\left(  \sin\left(  \frac{1}{1+x^{2}}\right)  -x+\left(
1+c\right)  t,t+\frac{1}{1+x^{2}}\right)  $ with $t$ as the animation
parameter: a soliton made up of non-soliton displacements.}
\label{1D2Dfig}
\end{center}
\end{figure}

In order to explicitly demonstrate how one can construct solitons of this
sort, let $k=0$, $v=\frac{1}{1+x^{2}}$, and $u=\sin\left(  \frac{1}{1+x^{2}
}\right)  -2x+ct$. Now, Equation \ref{1D2Deq} is satisfied and the graph
$\left(  x+u,v\right)  $ shown in Figure \ref{1D2Dfig} is that of a slightly
slanted bell-shaped soliton moving towards right (left) with a velocity of
$c>0$ ($c<0)$. This is an example of a soliton motion resulting from
non-soliton displacement functions.

If only transverse motion is allowed, i.e. $u=0$ and $k=0$, as in the case of
classical transverse motion of a vibrating string, then one gets
\begin{equation}
v_{t}=-cv_{x}
\end{equation}
which is the equation for a classical soliton, leading to d'Alembert's solutions.

\begin{figure}
[ht]
\begin{center}
\fbox{\includegraphics[
height=2.2684in,
width=3.403in
]
{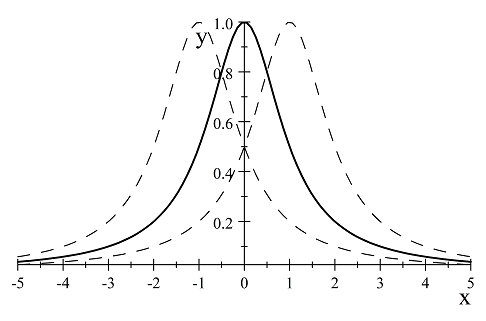}
}\caption{A bell-shaped soliton that moves back and forth in $x$ direction
between the dashed boundaries.}
\label{rocks}
\end{center}
\end{figure}

As an example for unusual soliton motions, we now consider a soliton motion
for which $k=0$ and $u=0$, but $c=\sin t$. Thus,
\begin{equation}
v_{t}+\left(  \sin t\right)  v_{x}=0
\end{equation}
the solution of which is
\begin{equation}
v=f\left(  x+\cos t\right)
\end{equation}
where $f$ is a continuously differentiable function with respect its argument.
As an example, one can plot $\left(  x,\frac{1}{1+\left(  x+\cos t\right)
^{2}}\right)  $ using $t$ as the animation parameter. The result, shown in
Figure \ref{rocks}, is a bell-shaped curve that rocks back and forth in $x$ direction.

\subsection{2D Membrane in 3D Motion}

In this case, $n=2$, $m=1$. By taking $k=0$ the soliton condition becomes
\begin{equation}
v_{t}=\left[
\begin{array}
[c]{cc}
v_{x} & v_{y}
\end{array}
\right]  \left[
\begin{array}
[c]{cc}
1+u_{1,x} & u_{1,y}\\
u_{2,x} & 1+u_{2,y}
\end{array}
\right]  ^{-1}\left[
\begin{array}
[c]{c}
u_{1,t}-c_{1}\\
u_{2,t}-c_{2}
\end{array}
\right]
\end{equation}
where $c_{i}$ are the velocity components of the graph in $xy$-plane. Numerous
interesting cases can be obtained from this relation. For example, for
$u_{1}=f_{1}\left(  x-c_{1}t\right)  $ and $u_{2}=f_{2}\left(  y-c_{2}
t\right)  $ the kinematic condition is met by $v=g_{1}\left(  x-c_{1}t\right)
+g_{2}\left(  y-c_{2}t\right)  $, which render the graph $\left(
x+u_{1},y+u_{2},v\right)  $ a soliton, a fact that we checked using a graph
animation software. This can be easily extended to higher dimensions.
Nevertheless, more complicated situations make up the dominating class.

If the motion is restricted to transverse directions only, this condition
reduces to
\begin{equation}
v_{t}=-c_{1}v_{x}-c_{2}v_{y} \label{2D3Deq}
\end{equation}
which is simply an extension of the previous case and a 2D version of the
transverse wave solitons presented before. If $c_{i}$ are functions of time,
then the general solution is obtained by using the method of characteristic:
\begin{equation}
v\left(  x,y,t\right)  =f\left(  x-D_{1}\left(  t\right)  ,y-D_{2}\left(
t\right)  \right)
\end{equation}
where $f$ is function of two variables and $\frac{dD_{i}}{dt}=c_{i}\left(
t\right)  $ are the velocity components. For example, by letting
$v=e^{-\left(  x-\cos t\right)  ^{2}}e^{-\left(  y-\sin t\right)  ^{2}}$, the
graph $\left(  x,y,e^{-\left(  x-\cos t\right)  ^{2}}e^{-\left(  y-\sin
t\right)  ^{2}}\right)  $ becomes a lump moving on a circle centered at
origin, which is verified by using a graph animation software. This is an
example for a soliton moving with a varying velocity direction.

It is possible to construct quite arbitrarily moving solitons using this
result. Below, we give a simple demonstration of a lump in a 2D membrane using
cylindrical coordinates, and then show that the displacement components
satisfy the soliton condition.

\begin{figure}
[ht]
\begin{center}
\includegraphics[
height=1.7608in,
width=3.5189in
]
{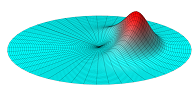}
\caption{A 2D membrane moving in 3D. Motion of the points are in the lateral
direction whereas the lump travels on a circle around the origin.}
\label{membraneLumpd}
\end{center}
\end{figure}

Figure \ref{membraneLumpd} shows a membrane, points of which move in the
transverse direction. The lump in the membrane moves on a circle in membrane
plane, centered at origin.

The graph is obtained by plotting $\left(  r,\theta,\frac{1}{2}e^{-\left(
\frac{\theta-\omega t}{\alpha}\right)  ^{2}}e^{-\left(  \frac{r-2}{\beta
}\right)  ^{2}}\right)  $ in cylindrical coordinates, where $t$ is the
animation parameter corresponding to time. Here, $\omega$ is the angular
velocity of the lump, and, the parameters $\alpha$ and $\beta$ are arbitrary
scaling factors that shape the lump.

The displacements are given by
\begin{align}
u\left(  r,\theta,t\right)   &  =0\\
v\left(  r,\theta,t\right)   &  =\frac{1}{2}e^{-\left(  \frac{\theta-\omega
t}{\alpha}\right)  ^{2}}e^{-\left(  \frac{r-2}{\beta}\right)  ^{2}}
\end{align}
from which one calculates the derivatives involved in the soliton condition as
follows.
\begin{align}
v_{t}  &  =2\omega\left(  \frac{\theta-\omega t}{\alpha}\right)  v\\
v_{x}  &  =2\left(  \frac{1}{r}\left(  \frac{\theta-\omega t}{\alpha}\right)
\sin\theta-\left(  \frac{r-2}{\beta}\right)  \cos\theta\right)  v\\
v_{y}  &  =2\left(  -\frac{1}{r}\left(  \frac{\theta-\omega t}{\alpha}\right)
\cos\theta-\left(  \frac{r-2}{\beta}\right)  \sin\theta\right)  v
\end{align}
When these are inserted in Equation \ref{2D3Deq}, the result is
\begin{align}
\omega\left(  \frac{\theta-\omega t}{\alpha}\right)   &  =\frac{1}{r}\left(
-c_{1}\sin\theta+c_{2}\cos\theta\right)  \left(  \frac{\theta-\omega t}
{\alpha}\right) \nonumber\\
&  +\left(  c_{1}\cos\theta+c_{2}\sin\theta\right)  \left(  \frac{r-2}{\beta
}\right)
\end{align}
Since $\frac{\theta-\omega t}{\alpha}$ and $\frac{r-2}{\beta}$ are
independent, one must have
\begin{align}
-c_{1}\sin\theta+c_{2}\cos\theta &  =\omega r\\
c_{1}\cos\theta+c_{2}\sin\theta &  =0
\end{align}
the solution of which is
\begin{equation}
\left[
\begin{array}
[c]{c}
c_{1}\\
c_{2}
\end{array}
\right]  =\omega r\left[
\begin{array}
[c]{c}
\sin\theta\\
-\cos\theta
\end{array}
\right]  =\omega\left[
\begin{array}
[c]{c}
y\\
-x
\end{array}
\right]
\end{equation}
Clearly, these velocity components correspond to a velocity that is tangent to
a circle centered at the origin, with a speed of $\omega r$, or a constant
angular speed of $\omega$. It is also possible to design a soliton moving on a
circle with a variable speed. Note that this is an example for a soliton
velocity that depends on spatial coordinates. Yet, the shape of the soliton is conserved.

\subsection{1D String in (M+1)D Motion}

In this case $n=1$, $m=M\geqslant1$. For $\bar{k}=\bar{0}$ the kinematic
condition becomes
\begin{align}
\bar{v}_{t}  &  =\left(  \frac{u_{t}-c}{1+u_{x}}\right)  \bar{v}_{x}\\
v_{i,t}+cv_{i,x}  &  =u_{t}v_{i,x}-u_{x}v_{i,t} \label{1DMDString}
\end{align}
for all $i=1,...,M$. For $M=1$ this reduces to the case presented as 1D string
moving in 2D.

\textbf{Case 1}: $u_{t}-c=0$. In this case, $v_{i,t}=0$, i.e. $v_{i}
=f_{i}\left(  x\right)  $, for all $i$. Further, $u=ct+g\left(  x\right)  $
and the apparent motion of the graph $\left(  x+ct+g(x),f_{1}\left(  x\right)
,\cdots\right)  $ would be a soliton with velocity $c$ in $x$ dimension. This
is verified by using a graph animation software.

\textbf{Case 2}: $u_{t}-c\neq0$. In this case, one has
\begin{equation}
v_{i,t}v_{j,x}=v_{i,x}v_{j,t}
\end{equation}
for any pair of $i$ and $j$. That is, regardless of $u$, the transverse
displacement functions must be compatible via these equations. In order to see
what this amounts to we shall investigate all possible sub-cases.

2a) If for any $i$, $v_{i,t}=0$, then $v_{i,x}=0$, i.e. $v_{i}\left(
x,t\right)  =c_{1}$, a constant. This means the whole base manifold shifts in
$v_{i}$ direction by $c_{1}$, which is not an interesting case.

2b) If for a particular $i$, $v_{i,x}=0$, i.e. $v_{i}=f_{i}\left(  t\right)
$, and $v_{i,t}$ is not identically zero, then the solutions for all other
functions are also in the form $v_{j}=f_{j}\left(  t\right)  $. In this case,
$u=-x+g\left(  t\right)  $. This is rejected since all of the base manifold is
mapped to a single point at any given time.

2c) If $\frac{v_{i,t}}{v_{i,x}}=f\left(  x,t\right)  $ is a general non-zero
function of $x$ and $t$, then
\begin{align}
\frac{v_{j,t}}{v_{j,x}}  &  =f\left(  x,t\right)  \text{ for all }j\\
\frac{u_{t}-c}{1+u_{x}}  &  =f\left(  x,t\right)
\end{align}
Letting $u=U-x+ct$ yields
\begin{equation}
\frac{U_{t}}{U_{x}}=f\left(  x,t\right)
\end{equation}
Thus, all $v_{i}$ and $U$ satisfy the same equation.

For example, if $f\left(  x,t\right)  =-c_{1}$, a constant, then
$v_{i}=g\left(  x-c_{1}t\right)  $ and $U=h\left(  x-c_{1}t\right)  $ become
solitons, in their own right, with a velocity of $c_{1}$. Then, $u=h\left(
x-c_{1}t\right)  -x+ct$ or $u=H\left(  x-c_{1}t\right)  +\left(
c-c_{1}\right)  t $. It is interesting to note that for $c_{1}=c$, all
motions, including that of the graph, become solitons with the same velocity.
However, for $c_{1}=-c$, $v_{i}$ and $U$ are solitons with a velocity opposite
of that of the graph. In the latter case, $u$ is the sum of two solitons
moving with equal and opposite velocities. More on this is presented in the
following subsection.

\subsection{All Solitons}

A special case is when all motions are solitons, including the graph. Thus, we
take $\bar{u}_{t}=-\left(  \bar{\nabla}\bar{u}^{T}\right)  ^{T}\bar{c}_{u}$
and $\bar{v}_{t}=-\left[  \bar{\nabla}\bar{v}^{T}\right]  ^{T}\bar{c}_{v}$,
and enforce the graph soliton condition as follows.
\begin{equation}
-\left[  \bar{\nabla}\bar{v}^{T}\right]  ^{T}\bar{c}_{v}=\left[  \bar{\nabla
}\bar{v}^{T}\right]  ^{T}\left[  \tilde{I}+\left(  \bar{\nabla}\bar{u}
^{T}\right)  ^{T}\right]  ^{-1}\left(  -\left(  \bar{\nabla}\bar{u}
^{T}\right)  ^{T}\bar{c}_{u}-\bar{c}\right)
\end{equation}
\begin{equation}
\left[  \bar{\nabla}\bar{v}^{T}\right]  ^{T}\left[  \left[  \tilde{I}+\left(
\bar{\nabla}\bar{u}^{T}\right)  ^{T}\right]  ^{-1}\left(  -\left(  \bar
{\nabla}\bar{u}^{T}\right)  ^{T}\bar{c}_{u}-\bar{c}\right)  +\bar{c}
_{v}\right]  =\bar{0}
\end{equation}
The non-trivial solutions require the vector inside the brackets to be in the
null space of $\left[  \bar{\nabla}\bar{v}^{T}\right]  ^{T}$. Investigating
such a situation is quite involved and outside the scope of this study.
Further, in many cases the rank of the $m\times n$ matrix $\left[  \bar
{\nabla}\bar{v}^{T}\right]  ^{T}$ is actually $n$, giving an empty null space.
Therefore, we concentrate on the trivial solutions
\begin{equation}
\left[  \tilde{I}+\left(  \bar{\nabla}\bar{u}^{T}\right)  ^{T}\right]
^{-1}\left(  -\left(  \bar{\nabla}\bar{u}^{T}\right)  ^{T}\bar{c}_{u}-\bar
{c}\right)  +\bar{c}_{v}=\bar{0}
\end{equation}
which, after manipulations, yields the following.
\begin{equation}
\left[  \tilde{I}+\left(  \bar{\nabla}\bar{u}^{T}\right)  ^{T}\right]  \left(
\bar{c}_{v}-\bar{c}_{u}\right)  =\bar{c}-\bar{c}_{u} \label{trivialEq}
\end{equation}
Each row of this equation is a linear PDE involving only $u_{i}$. Given
$\bar{u}$ and any two velocities, one can solve the unknown velocity. Or,
given the velocities one may look for solutions for $\bar{u}$ that are
solitons, which may or may not exist. All such combinations may open up an
interesting avenue for further research. However, this is not the aim of the
current study. Again, we only consider the trivial cases and their
implications. The following corollary follows from the invertibility of
$\left(  \tilde{I}+\left(  \bar{\nabla}\bar{u}^{T}\right)  ^{T}\right)  $ and
Equation \ref{trivialEq}.

\begin{corollary}
\label{CorAllSol}Given that $u$, $v$, and $\left(  x+u,v\right)  $ are
solitons with velocities $\bar{c}_{u}$, $\bar{c}_{v}$, and $\bar{c}$,
respectively, then $\bar{c}_{v}=\bar{c}_{u}=\bar{c}$ whenever $\bar{c}
_{u}=\bar{c}_{v}$ or $\bar{c}_{u}=\bar{c}$.
\end{corollary}

Note that for $\bar{c}_{v}=\bar{c}$ case, $\left(  \bar{c}_{v}-\bar{c}
_{u}\right)  $ becomes an eigenvector of $\tilde{I}+\left(  \bar{\nabla}
\bar{u}^{T}\right)  ^{T}$ corresponding to an eigenvalue of 1, if exists. An
example is provided by $u_{1}=\left(  x-t\right)  +\left(  y-t\right)  $ and
$u_{2}=x+t+\left(  y+t\right)  $, with $\bar{c}_{u}=\left[
\begin{array}
[c]{cc}
1 & -1
\end{array}
\right]  ^{T}$. These give $\tilde{I}+\left(  \bar{\nabla}\bar{u}^{T}\right)
^{T}=\left[
\begin{array}
[c]{cc}
2 & 1\\
1 & 2
\end{array}
\right]  $ that has an eigenvector of $\left[
\begin{array}
[c]{cc}
-1 & 1
\end{array}
\right]  ^{T}$ with a unit eigenvalue. Therefore, the soliton condition is met
if
\begin{equation}
\bar{c}_{v}=\bar{c}=\bar{c}_{u}+\alpha\left[
\begin{array}
[c]{cc}
-1 & 1
\end{array}
\right]  ^{T}=\alpha^{\ast}\bar{c}_{u}
\end{equation}
where $\alpha^{\ast}\in R$.

A stronger condition is the following.

\begin{corollary}
\label{Cor3}If $\bar{u}$ and the graph $\left(  \bar{x}+\bar{u},\bar
{v}\right)  $ are solitons of velocity\ $\bar{c}$ then $\bar{v}$ is a soliton
of velocity $\bar{c}$, too.
\end{corollary}

\begin{proof}
In this case, $\bar{u}_{t}=-\left(  \bar{\nabla}\bar{u}^{T}\right)  ^{T}
\bar{c} $ and the soliton condition becomes
\begin{align}
\bar{v}_{t}  &  =\left[  \bar{\nabla}\bar{v}^{T}\right]  ^{T}\left[  \tilde
{I}+\left(  \bar{\nabla}\bar{u}^{T}\right)  ^{T}\right]  ^{-1}\left(  -\left(
\bar{\nabla}\bar{u}^{T}\right)  ^{T}\bar{c}-\bar{c}\right) \\
\bar{v}_{t}  &  =-\left[  \bar{\nabla}\bar{v}^{T}\right]  ^{T}\bar{c}
\end{align}
indicating that $\bar{v}$ is a soliton of velocity $\bar{c}$.
\end{proof}

Another special case, as an extension of previous section, is when $\bar{v}$
is a vector soliton. In such a case, one would have $\bar{v}_{t}=-\left(
\bar{\nabla}\bar{v}^{T}\right)  ^{T}\bar{c}_{v}$. Then, by Theorem
\ref{MainTheo}
\begin{equation}
\left[  \bar{\nabla}\bar{v}^{T}\right]  ^{T}\left[  \left[  \tilde{I}+\left(
\bar{\nabla}\bar{u}^{T}\right)  ^{T}\right]  ^{-1}\left(  \bar{u}_{t}-\bar
{c}\right)  +\bar{c}_{v}\right]  =0\label{EqAllSol}
\end{equation}
Again, we concentrate on the trivial solutions and define $\bar{u}=\bar
{U}-\left(  \bar{x}-\bar{c}t\right)  $, which results in
\begin{equation}
\bar{U}_{t}=-\left(  \bar{\nabla}\bar{U}^{T}\right)  ^{T}\bar{c}_{v}
\end{equation}
Hence, $\bar{U}$ is a vector soliton with the same velocity as that of
$\bar{v}$. Then,
\begin{equation}
\bar{u}=\bar{f}\left(  \bar{x}-\bar{c}_{v}t\right)  -\left(  \bar{x}-\bar
{c}t\right)
\end{equation}
Note that if $\bar{c}=\bar{c}_{v}$ then $\bar{u}$ is a vector soliton with
velocity $\bar{c}$, and vice versa. This proves the following.

\begin{corollary}
\label{Cor1}If $\bar{v}$ is a soliton of velocity $\bar{c}_{v}$ then the graph
$\left(  \bar{x}+\bar{u},\bar{v}\right)  $ is a soliton of velocity\ $\bar{c}$
if either

\begin{enumerate}
\item $\bar{c}_{v}=\bar{c}$ and $\bar{u}$ is a soliton with velocity $\bar{c}
$ (Corollary \ref{CorAllSol}), or

\item $\bar{u}$ is the sum of two soliton vectors $\bar{f}\left(  \bar{x}
-\bar{c}_{v}t\right)  $ and $-\left(  \bar{x}-\bar{c}t\right)  $.
\end{enumerate}
\end{corollary}

Particular examples of this are presented in following sections. Note that
corollaries \ref{CorAllSol}, \ref{Cor3}, and \ref{Cor1} do not claim that the
displacement vectors have to be solitons for the graph to be a soliton.
Examples to the contrary were presented previously. What is important is that
they pave the way for the possibility of having soliton motions for all
functions involved.

\begin{figure}
[ht]
\begin{center}
\fbox{\includegraphics[
height=2.0954in,
width=2.1525in
]
{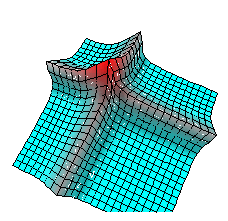}
}\caption{Two moving ridges obtained by plotting $\left(  x+e^{-\left(
x-t\right)  ^{2}},y+e^{-\left(  y-t\right)  ^{2}},\frac{1}{1+\left(
x-t\right)  ^{2}}+\frac{1}{1+\left(  y-t\right)  ^{2}}\right)  $}
\label{Ridges}
\end{center}
\end{figure}

\begin{figure}
[ht]
\begin{center}
\fbox{\includegraphics[
height=1.6898in,
width=2.1465in
]
{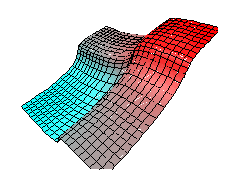}
}\caption{Two moving kinks obtained by plotting $\left(  x+e^{-\left(
x-t\right)  ^{2}},y+e^{-\left(  y-t\right)  ^{2}},\arctan\left(  x-t\right)
+\arctan\left(  y-t\right)  \right)  $}
\label{TwoKinks}
\end{center}
\end{figure}

As a demonstration one can take $u_{1}=f_{1}\left(  x-c_{1}t\right)  $,
$u_{2}=f_{2}\left(  y-c_{2}t\right)  $, and $v=g_{1}\left(  x-c_{1}t\right)
+g_{2}\left(  y-c_{2}t\right)  $, a 2D membrane moving in 3D. The plot
$\left(  x+u_{1},y+u_{2},v\right)  $ is a soliton, as checked in a graph
animation software. This is different from the membrane example given earlier,
in which the motion was restricted to the transverse direction. Now, we have
motions in all directions. The figures below demonstrate two examples. Figure
\ref{Ridges} shows two ridges in cross formation moving along a 45-degree line
in $xy$-plane. Figure \ref{TwoKinks} shows two smooth kinks moving similarly.

\subsection{Standing Waves}

For a standing wave we set $\bar{C}=\bar{0}$. Then, the soliton condition
becomes
\begin{equation}
\bar{v}_{t}=\left[  \bar{\nabla}\bar{v}^{T}\right]  ^{T}\left[  \tilde
{I}+\left(  \bar{\nabla}\bar{u}^{T}\right)  ^{T}\right]  ^{-1}\bar{u}_{t}
\end{equation}

For $R\times R$ case, one can show that the solutions are of the form
$v=f\left(  x+u\right)  $, where all the functions involved are arbitrary,
except for the condition of differentiability. The resulting graph $\left(
x+u,f\left(  x+u\right)  \right)  $ is a standing wave regardless of $u$. For
example, for $u=\sin x+\cos t$ and $v=e^{-\left(  x+u\right)  ^{2}}$one gets a
graph
\begin{equation}
\left(  x+\sin x+\cos t,e^{-\left(  x+\sin x+\cos t\right)  ^{2}}\right)
\end{equation}
of a standing bell-shaped curve, the peak of which occurs at $x=0$, despite
the fact that the material points are moving with non-zero velocities.

Also, if $\left[  \tilde{I}+\left(  \bar{\nabla}\bar{u}^{T}\right)
^{T}\right]  ^{-1}\bar{u}_{t}=-\bar{c}$ then $\bar{v}_{t}=-\left[  \bar
{\nabla}\bar{v}^{T}\right]  ^{T}\bar{c}$, hence, $\bar{v}$ is a soliton with
velocity $\bar{c}$. Further,
\begin{equation}
\bar{u}_{t}=-\left[  \tilde{I}+\left(  \bar{\nabla}\bar{u}^{T}\right)
^{T}\right]  \bar{c}
\end{equation}
Letting $\bar{u}=\bar{U}-\bar{c}t$ yields $\bar{U}_{t}=-\left(  \bar{\nabla
}\bar{u}^{T}\right)  ^{T}\bar{c}$, a soliton. Hence, the following is proven.

\begin{corollary}
If $\bar{v}$ is a soliton of velocity $\bar{c}$ and $\bar{u}=\bar{U}-\bar{c}t
$, where $\bar{U}$ is a soliton of velocity $\bar{c}$, then the graph is a
standing wave.
\end{corollary}

\subsection{Traveling Knot}

\begin{center}

\begin{figure}
[b]
\begin{center}
\includegraphics[
height=1.6821in,
width=3.0338in
]
{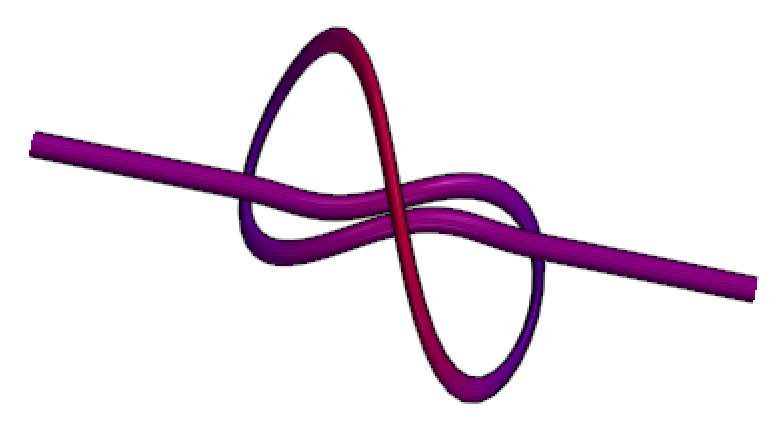}
\caption{A soliton knot.}
\label{solKnot}
\end{center}
\end{figure}

\end{center}

A most interesting example involves the possibility of a knot moving in a 3D
manifold. In order to achieve this, we used the following Cartesian graph as
an example.
\begin{equation}
\left(  x-\frac{5z}{1+z^{4}},e^{-z^{2}}\sin\left(  4z\right)  ,e^{-z^{2}}
\cos\left(  4z\right)  \right)  \label{KnotEqu}
\end{equation}
where $z=x-t$. Note that since $u$ and $\bar{v}$ are solitons with velocity
$+1$, then the graph is necessarily a soliton of velocity $+1$, due to
Corollary \ref{CorAllSol}. In Figure \ref{solKnot}, the shape of the knot at
$t=0$ is shown. The knot moves towards right with a constant velocity as
confirmed by a graph animation software.

Note that, in this case, all displacement components are solitons with the
same speeds. Therefore, they all obey the classical wave equation:
$f_{tt}=c^{2}f_{xx}$. This exemplifies the fact that, with compatible initial
conditions, a soliton knot is plausible provided that compatible equations of
motion exist, i.e. the classical wave equations. In such a situation, if the
initial shape is a knot with appropriate velocities, then the ensuing motion
will be a knot with a preserved shape: a soliton knot.

This example also demonstrates the usefulness of Theorem \ref{MainTheo}.
Transforming Equation \ref{KnotEqu} into an explicit soliton form such as
$\left(  p,f\left(  p-t\right)  ,g\left(  p-t\right)  \right)  $, which would
serve as a direct demonstration of soliton character, seems impossible. Yet,
Theorem \ref{MainTheo} guarantees that the soliton condition is met since
equations \ref{1DMDString} are satisfied.

\section{Compatibility of Equations of Motions}

The kinematic condition for soliton motions, as presented in this study, is of
purely geometric character. The physics of the continuum, on the other hand,
will have certain equations of motions that any allowable motion would have to
satisfy. Therefore, for any soliton motion of a continuum with given physical
rules, equations of motion, one would have two sets of equations to be
satisfied by candidate motions.

By compatibility of the equations of motions with the soliton condition we
mean that the set of solutions of the equations of motions that also satisfy
the soliton condition is not empty. If a system of equations of motion do not
admit any solutions that also satisfy the soliton condition, then the
underlying physics is not compatible. In such a case, if one insists on having
soliton solutions then the physical model will have to be modified. On the
other hand, if one insists on the physics, then there would be no soliton solutions.

For example, not all physical models will admit a traveling knot solution.
However, if a traveling knot solution is possible, then there must be certain
physical laws that govern and are compatible.

We shall now apply this to a 1D string in 2D motion. Let the physics be such
that both the transverse and longitudinal displacements are as described by
the classical wave equation. That is
\begin{equation}
v_{tt}=c_{v}^{2}v_{xx}\text{ \ and \ }u_{tt}=c_{u}^{2}u_{xx}
\end{equation}
We also insist that the graph $\left(  x+u,v\right)  $ be a soliton with a
velocity of $c$ in $x$-direction. Hence, the soliton condition becomes
$v_{t}+cv_{x}=v_{x}u_{t}-u_{x}v_{t}$, as shown before. Disregarding the
boundary conditions, the general forms of the solutions to the equations of
motion are
\begin{align}
v\left(  x,t\right)   &  =f\left(  x-c_{v}t\right) \\
u\left(  x,t\right)   &  =g\left(  x-c_{u}t\right)
\end{align}
which, when used in Equation \ref{1D2Deq}, result in
\begin{equation}
-c_{v}f^{\prime}+cf^{\prime}=f^{\prime}\left(  -c_{u}g^{\prime}\right)
-\left(  -c_{v}f^{\prime}\right)  \left(  g^{\prime}\right)
\end{equation}
For $f^{\prime}\neq0$ this reduces to
\begin{equation}
c-c_{v}=\left(  c_{v}-c_{u}\right)  g^{\prime}
\end{equation}
from which one gets three cases as follows.

\textbf{Case 1}: $c=c_{v}\neq c_{u}$: In this case, one must have $g^{\prime
}=0$, meaning constant $u$. This constant shift can be taken as zero, which
turns the case into the classical transverse string motion and, $v$ and the
graph are arbitrary solitons with the same velocity.

\begin{figure}
[ht]
\begin{center}
\fbox{\includegraphics[
height=1.8109in,
width=2.7155in
]
{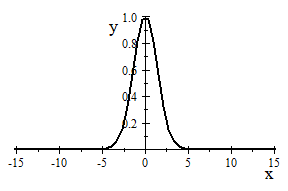}
}\caption{This graph looks like a classical soliton despite the fact that the
motion of points in $x$-direction is unbounded.}
\label{case2Fig}
\end{center}
\end{figure}

\textbf{Case 2}: $c\neq c_{v}\neq c_{u}$: Now, $u\left(  x,t\right)
=-\frac{c-c_{v}}{c_{u}-c_{v}}\left(  x-c_{u}t\right)  +d$, where $d$ is a
constant. Although, $v$ and the graph $\left(  x+u,v\right)  $ may still be
arbitrary and acceptable solitons, the displacement function $u$ is now
unbounded and, hence, physically not viable. An example is the following:
$\left(  x-\frac{3-2}{1-2}\left(  x-t\right)  ,e^{-\left(  x-2t\right)  ^{2}
}\right)  $, in which $u$ is a soliton with a velocity of +1, $v$ is a soliton
with a velocity of +2, and the graph is a soliton with a velocity of +3, see
Figure \ref{case2Fig}. The graph $\left(  2x-t,e^{-\left(  x-2t\right)  ^{2}
}\right)  $ is equivalent to $\left(  p,e^{-\left(  \frac{p-3t}{2}\right)
^{2}}\right)  $.

\begin{figure}
[ht]
\begin{center}
\fbox{\includegraphics[
height=1.721in,
width=2.5832in
]
{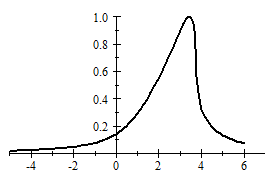}
}\caption{This graph and its underlying displacements are all solitons with
the same velocity.}
\label{Case3Fig}
\end{center}
\end{figure}

\textbf{Case 3}: $c_{u}=c_{v}=c$: In this case, $v$ and $u$, as well as the
graph, are arbitrary solitons with the same velocity, as predicted by
Corollary \ref{CorAllSol}. For example, for $u=e^{-\left(  x-t\right)  ^{2}}$
and $v=\frac{1}{1+\left(  x-t\right)  ^{2}}$, the graph of $\left(
x+e^{-\left(  x-t\right)  ^{2}},\frac{1}{1+\left(  x-t\right)  ^{2}}\right)  $
is a well-slanted, bell-shaped curve moving right in x-direction, as shown in
Figure \ref{Case3Fig}.

The final case is physically the most acceptable. Therefore, if the physics of
the displacements are governed by the classical wave equation, then a soliton
motion is plausible probably only if all the motions are solitons with the
same velocity.

\section{Conclusion}

This study shows that the motions of an $n$ dimensional body in $n+m$
dimensions admit soliton solutions if and only if the kinematic condition
described by Theorem \ref{MainTheo} is met, regardless of the underlying
physics. Special cases ranging from simple transverse waves to (M+1)D motions
of 1D strings, 3D motions of 2D membranes, and so on, are presented.
Plausibility of soliton knots based on physically acceptable wave motions are
demonstrated. It is shown that the case that all involved motions,
displacements and the graph, are solitons is admissible. Finally, the
compatibility of equations of motions with the kinematic condition is explored.

Given the equations of motion for a system, the presented kinematic condition
constrains the set of soliton solutions further, implications of which may be
significant. In simple cases implications seem to be not so strict, as was
shown in the case of 1D string moving in 2D governed by the classical wave
equation. However, in higher dimensional cases, or in cases involving more
complicated equations of motions, the problem may not be resolved in a
straightforward manner.

\end{document}